\documentclass[final,1p,times]{elsarticle} 
\usepackage{graphicx} 
\usepackage{amssymb} 
\usepackage{amsthm} 
 \usepackage{lineno} 
\usepackage{xspace}
\journal{Nuclear Physics A} 

\newcommand{\sqrts}{$\sqrt{s}$\xspace}
\newcommand{\pT}{$p_{T}$\xspace}
\newcommand{\pp}{{\it p-p}\xspace}

\begin{document} 

\begin{frontmatter} 


\title{Exploring Jet Properties in \pp Collisions at 200 GeV with STAR}

\author{Helen Caines for the STAR Collaboration}

\address{Physics Department, Yale University, New Haven, CT 06520, U.S.A}

\begin{abstract} 
The mechanisms underlying hadronization are not well understood, both in  vacuum and in hot QCD matter. Precise characterization of jet  fragmentation to hadrons in \pp collisions will help elucidate the fundamental process of hadronization, and will serve as essential  reference to measure the modification of hadronization in heavy ion collisions. We present measurements of fragmentation functions for  unidentified particles in jets produced  in \pp collisions at 200 GeV using the STAR detector at RHIC. The results from different jet reconstruction algorithms are compared,  including variations of the resolution parameter. It is found that the results are largely insensitive to details of the jet-finding algorithm at RHIC energies. Particle production  inside and outside of these reconstructed jets will be compared to improve our understanding of the hadronization mechanisms for soft and hard  particles in \pp events at RHIC energies. 

\end{abstract} 

\end{frontmatter} 

\section{Introduction and the Analysis}

\begin{figure}[htb]
	\begin{minipage}{0.46\linewidth}
		\begin{center}
			\includegraphics[width=\linewidth]{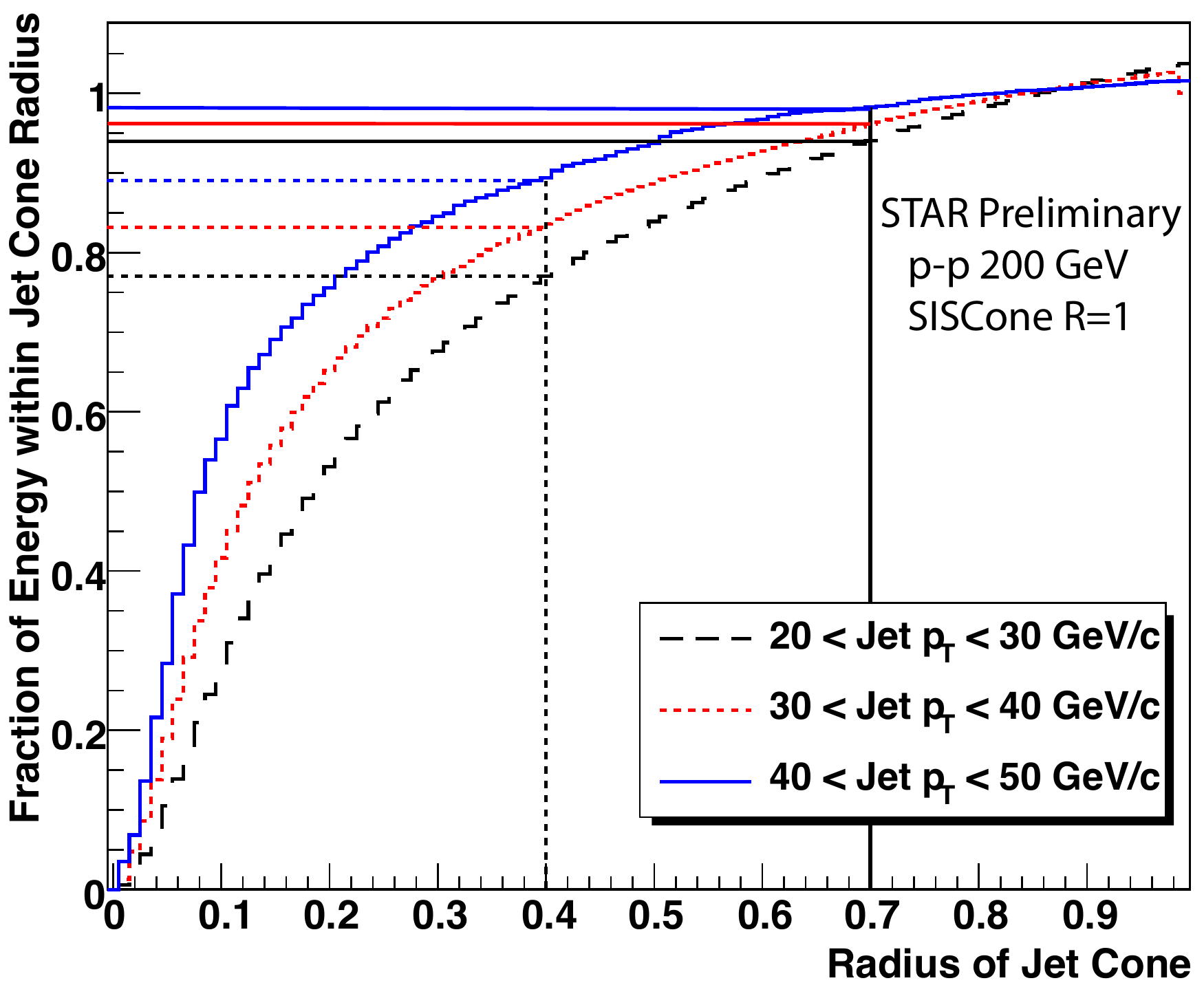}
		\end{center}
	\end{minipage}
	\hspace{1cm}
	\begin{minipage}{0.46\linewidth}
		\begin{center}
			\includegraphics[width=\linewidth]{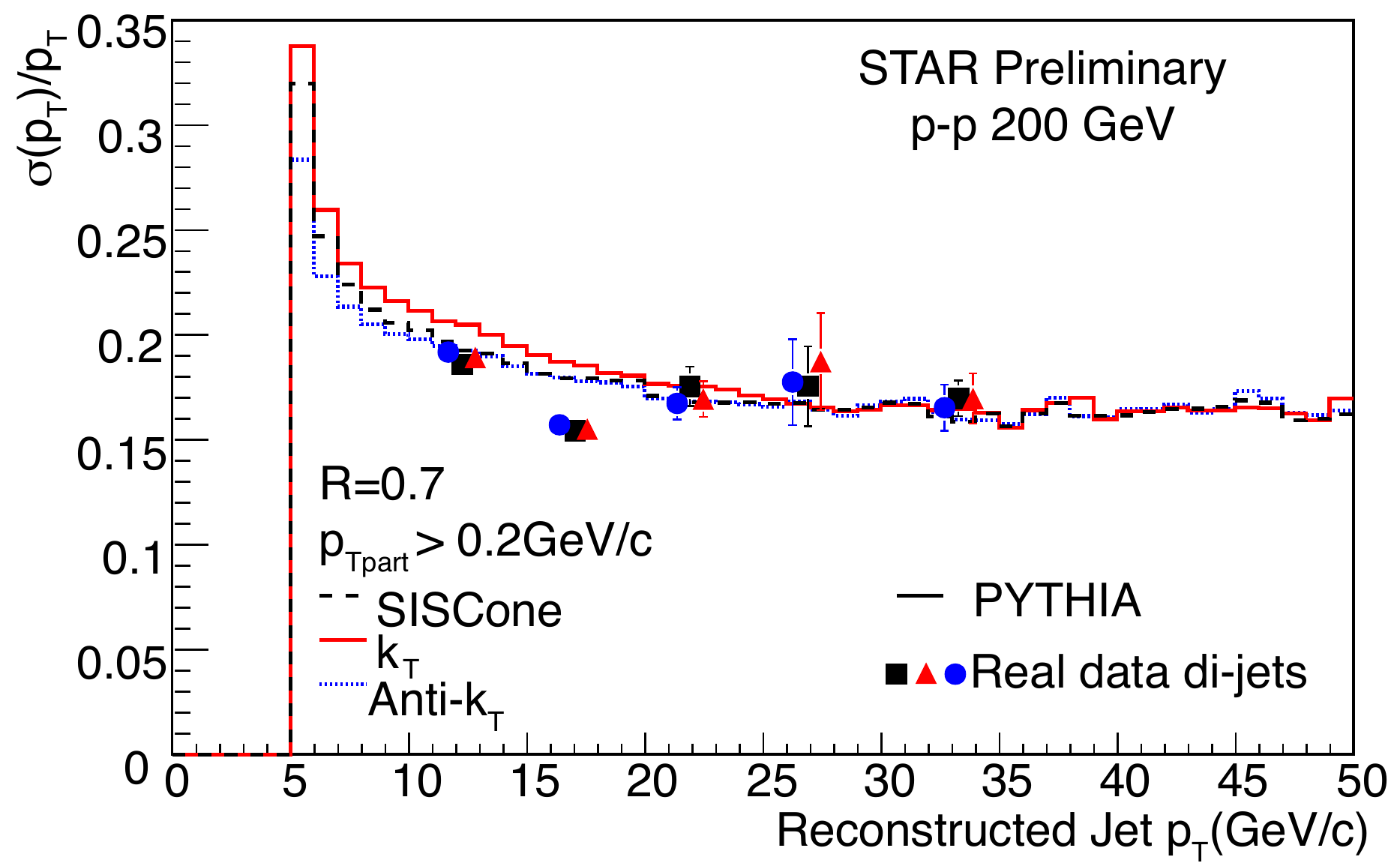}
		\end{center}

	\end{minipage}
	\caption{ Color online: Left: The fraction of the jet's energy contained within jet cone radius R. Jets were initially found using the SISCone algorithm with R=1.  Right: The reconstructed jet energy resolution determined from PYTHIA simulations (histograms) and real di-jet data for the three jet algorithms used (red triangles - k$_{T}$,  blue circles - anti-k$_{T}$ recombination and SISCone - black squares.	}
	\label{Fig:JetRes}
\end{figure}

The study of the properties of jets  and the underlying event in \pp collisions is important for improving our understanding of QCD and the hadronization process, as well as providing a vital baseline for comparison to measurements being performed in heavy-ion collisions~\cite{theseProc}. The results presented here are a preliminary study of  \pp collisions at \sqrts  = 200 GeV by the STAR collaboration from Run-6. We utilize the mid-rapidity Time Projection Chamber (TPC) and Barrel Electromagnetic Calorimeter (BEMC) to measure both the charged and neutral particle production. A jet-patch trigger, requiring E$_{T}>$8 GeV in a  $\Delta \eta$ x $\Delta \phi$ = 1x1 patch of the BEMC, was used to collect the data. This creates a neutral energy fragmentation bias for the triggering jet, hence charged particle fragmentation functions are presented only  for the di-jet partner not associated with the triggered jet-patch.  The k$_{T}$ and anti-k$_{T}$ recombination and SISCone jet algorithms from the FastJet package~\cite{fastjet} were used to reconstruct jets. A cut of $p_T$$>$0.2 GeV/c was applied to all charged particles considered in the event, and $E_{T} $$>$0.2 GeV for all neutral particles reconstructed in the BEMC. To investigate how the resolution parameters, R=$\sqrt{\Delta{\phi}^{2}+\Delta{\eta}^{2}}$,  used in this study affect the reconstructed jet energy, the SISCone algorithm was first run with R=1. The energy contained within the jet cone as a function of R was then studied, Fig.~\ref{Fig:JetRes}: Left panel. It can be seen that $>$75 (95)$\%$ of the jet's energy is contained within R=0.4 (0.7) for jets with \pT$>$ 20 GeV/c.  Figure ~\ref{Fig:JetRes}:Left panel also shows that higher energy jets are focussed within smaller jet radii.

The data are not yet corrected to the particle level and are therefore compared to PYTHIA 6.410~\cite{Pythia}, tuned to the CDF 1.96 TeV data (Tune A), predictions passed through STAR's simulations and reconstruction algorithms. There is a shift of the  reconstructed jet \pT  to lower values caused by detector inefficiences and undetected particles such as the neutron and K$^{0}_{L}$. The single particle reconstruction efficiency in the TPC is $>$ 80$\%$ for \pT$>$1 GeV/c.  The jet energy resolution was  obtained via two techniques. The first used the PYTHIA simulations to compare reconstructed jet energies at the particle and detector level. The second studied  the energy balance of ``back-to-back" di-jets in the real \pp data. Figure~\ref{Fig:JetRes}:Right panel shows that both methods resulted in  comparable jet energy resolutions of  $\sim$20$\%$ for reconstructed jet \pT $>$ 10 GeV/c.

\section{Jets in \pp}

\begin{figure}[htb]
	\begin{minipage}{0.46\linewidth}
		\begin{center}
			\includegraphics[width=\linewidth]{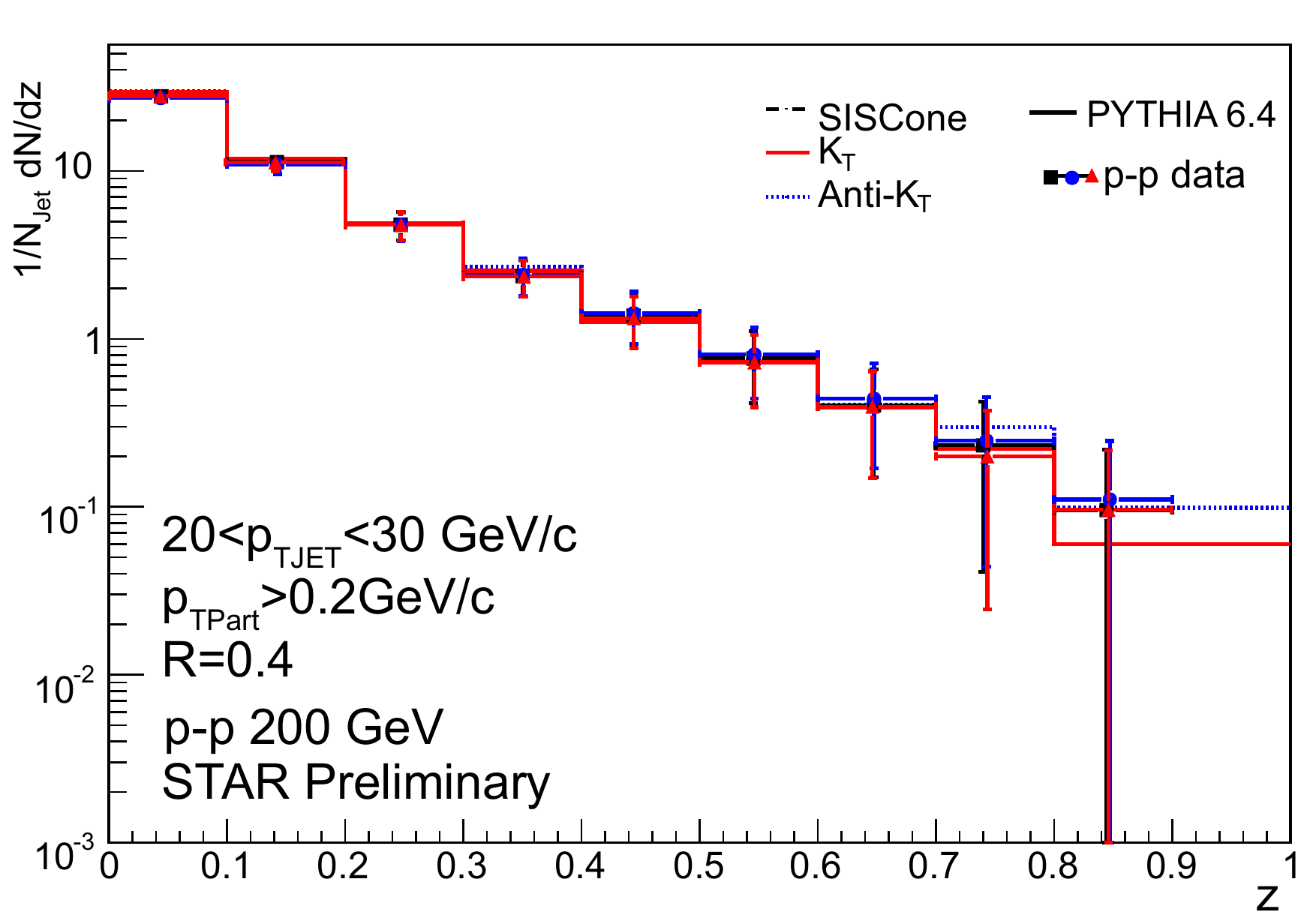}
		\end{center}
	\end{minipage}
	\hspace{1cm}
	\begin{minipage}{0.46\linewidth}
		\begin{center}
			\includegraphics[width=\linewidth]{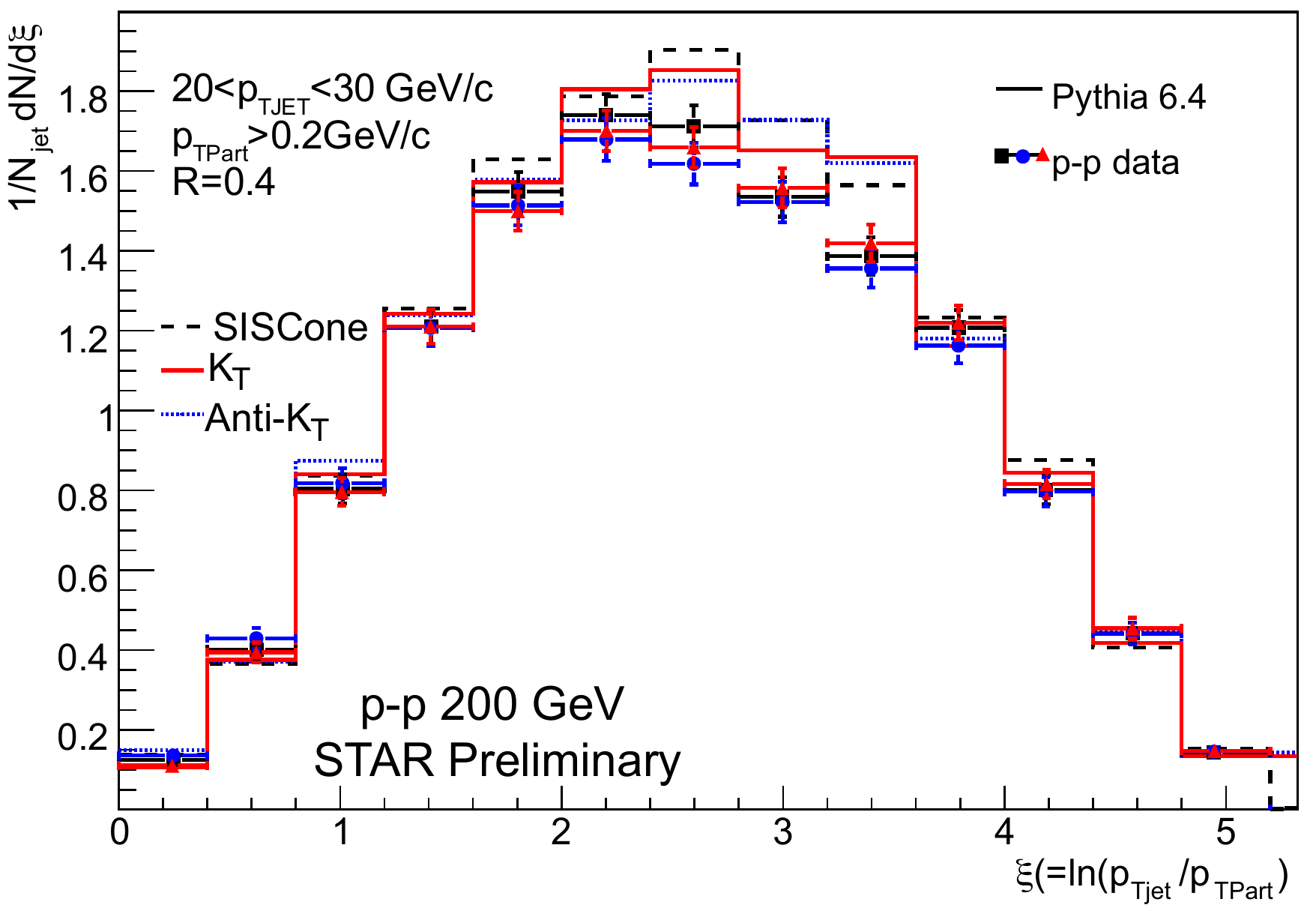}
		\end{center}

	\end{minipage}
	\caption{ Color online: Charged particle, detector level, $z$ and $\xi$ FF for jets reconstructed with 20$<$ \pT$<$ 30 GeV/c compared to PYTHIA for 3 different jet algorithms. $|\eta|<$1-R, R=0.4. Red triangles - k$_{T}$,  blue circles - anti-k$_{T}$ recombination and SISCone - black squares.}
	\label{Fig:FF04}
\end{figure}

\begin{figure}[htb]
	\begin{minipage}{0.46\linewidth}
		\begin{center}
			\includegraphics[width=\linewidth]{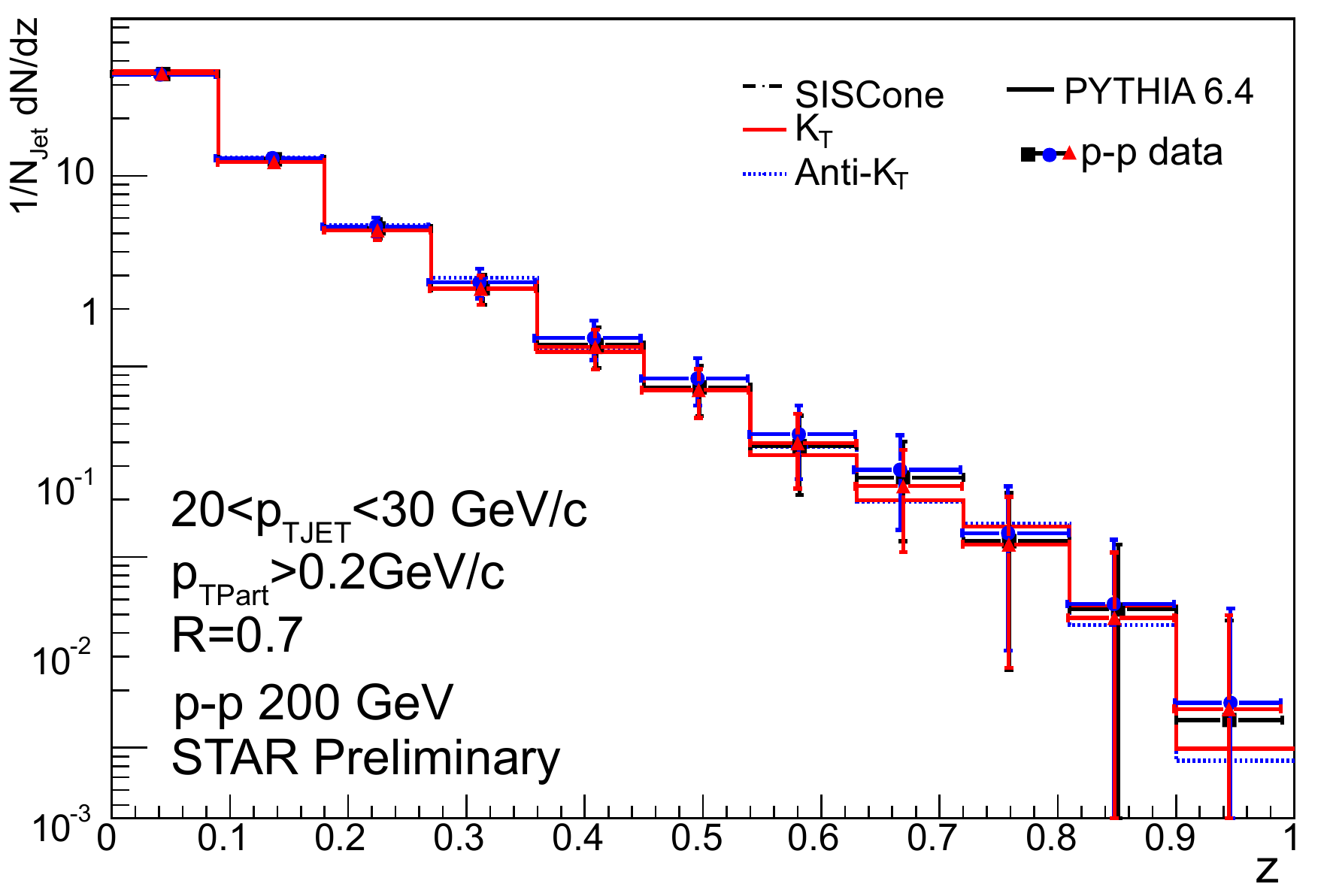}
		\end{center}
	\end{minipage}
	\hspace{1cm}
	\begin{minipage}{0.46\linewidth}
		\begin{center}
			\includegraphics[width=\linewidth]{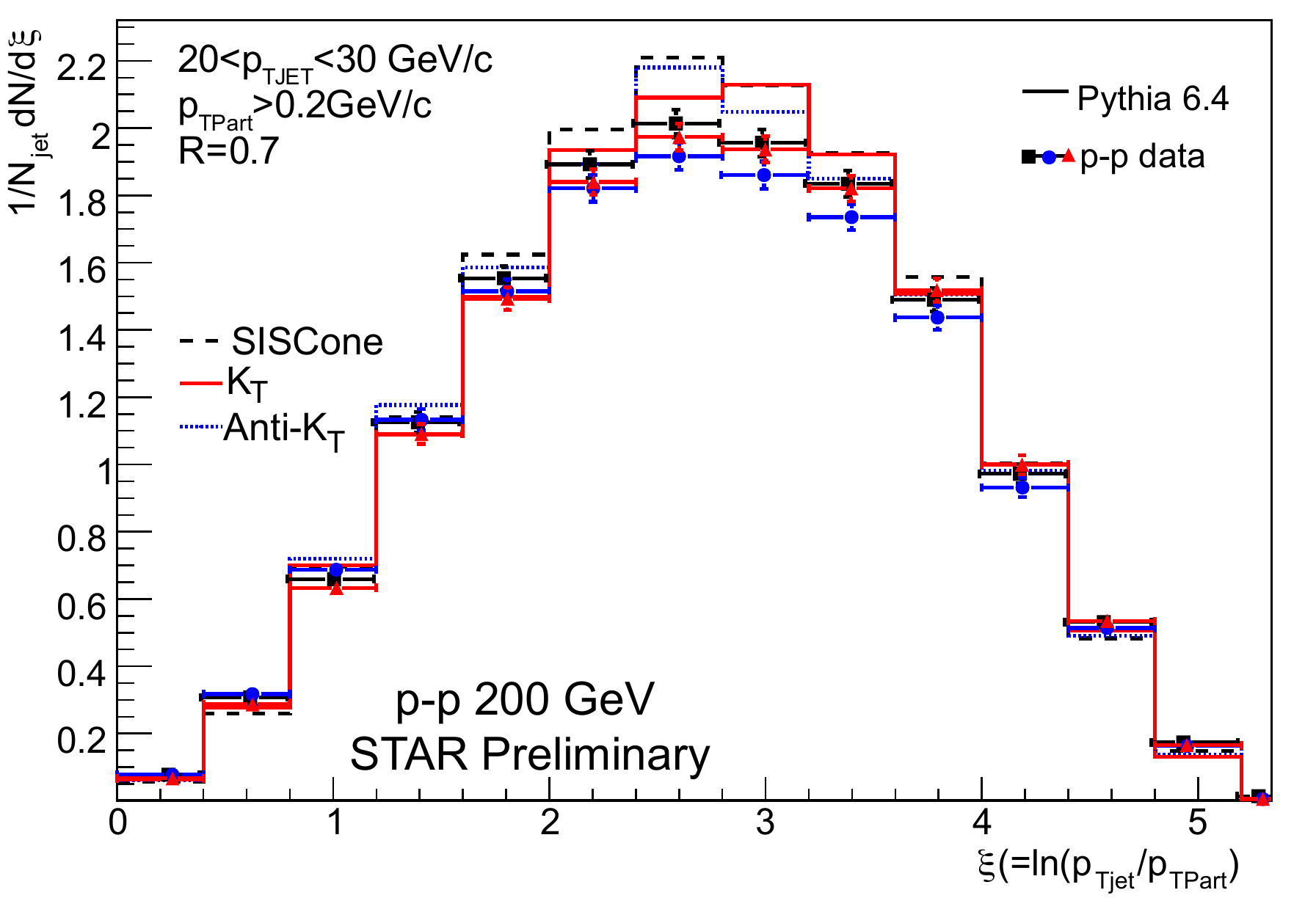}
		\end{center}
	\end{minipage}
	\caption{Color online: Charged particle, detector level, $z$ and $\xi$ FF for jets reconstructed with 20$<$ \pT$<$ 30 GeV/c compared to PYTHIA for 3 different jet algorithms. $|\eta|<$1-R, R=0.7. Red triangles - k$_{T}$,  blue circles - anti-k$_{T}$ recombination and SISCone - black squares.}
		\label{Fig:FF07}
\end{figure}

 The uncorrected charged particle fragmentation functions (FF) for jets with \pT reconstructed in the range 20-30 GeV/c are shown  for jet resolution parameters R=0.4, Fig.~\ref{Fig:FF04}, and R=0.7, Fig.~\ref{Fig:FF07}. The left plots are the FF as a function of z (=$p_T^{hadron}$/$p_T^{jet}$)  and right as a function of $\xi$(=ln(1/z)). The solid points are the data and the histograms are the PYTHIA simulations. There is reasonable agreement between the data and PYTHIA, and the different jet algorithms  (shown as different colors and line types/shapes in the figures) reconstruct the same FF within errors. This agreement, especially for the larger resolution parameter, suggests that there are only minor NLO contributions beyond those mimicked in the PYTHIA LO calculations at RHIC energies.

\section{The Underlying Event}

The Underlying Event (UE) in a \pp collision is defined as everything but the hard scattering. Thus, it has contributions from soft and semi-hard multiple parton interactions, initial and final state radiation and beam-beam remnants. Pile-up is not included as part of the UE. Since this study is performed at mid-rapidity the beam-beam contribution is minimal.  To study the UE we follow the CDF technique~\cite{CDF}. First the jets are reconstructed, next each event is split into four sections defined by their azimuthal  angle with respect to the leading jet axis ($\Delta{\phi}$). The range within $|\Delta{\phi}|$$<$60$^{0}$ is the lead jet region and an away jet area is designated for $|\Delta{\phi}|$$>$120$^{0}$. This leaves two transverse sectors of $60^{0}$$<$$\Delta{\phi}$$<$120$^{0}$ and $-120^{0}$$<$$\Delta{\phi}$$<$-60$^{0}$. One  is  called the TransMax region and is the transverse sector containing the largest charged particle multiplicity. The second sector is termed the TransMin region. Two analyses are then performed, a ``leading" jet study, where at least one jet is found in STAR's acceptance, and a ``back-to-back"   study which is a sub-set of the ``leading" jet collection.  This sub-set of events  has two (and only two) found jets  with $p_{T}^{awayjet}/p_{T}^{leadjet}$$>$0.7 and $|\Delta{\phi_{jet}}|$$>$150$^{0}$, this selection suppresses hard initial and final state radiation of the scattered parton. The TransMax region has an enhanced probability of containing contributions from these hard initial and final state radiation components. Thus, by comparing the TransMax and TransMin regions in the ``leading" and ``back-to-back" sets we can extract information about the various components in the UE.

\begin{figure}[ht]
		\begin{center}
			\includegraphics[width=0.6\linewidth]{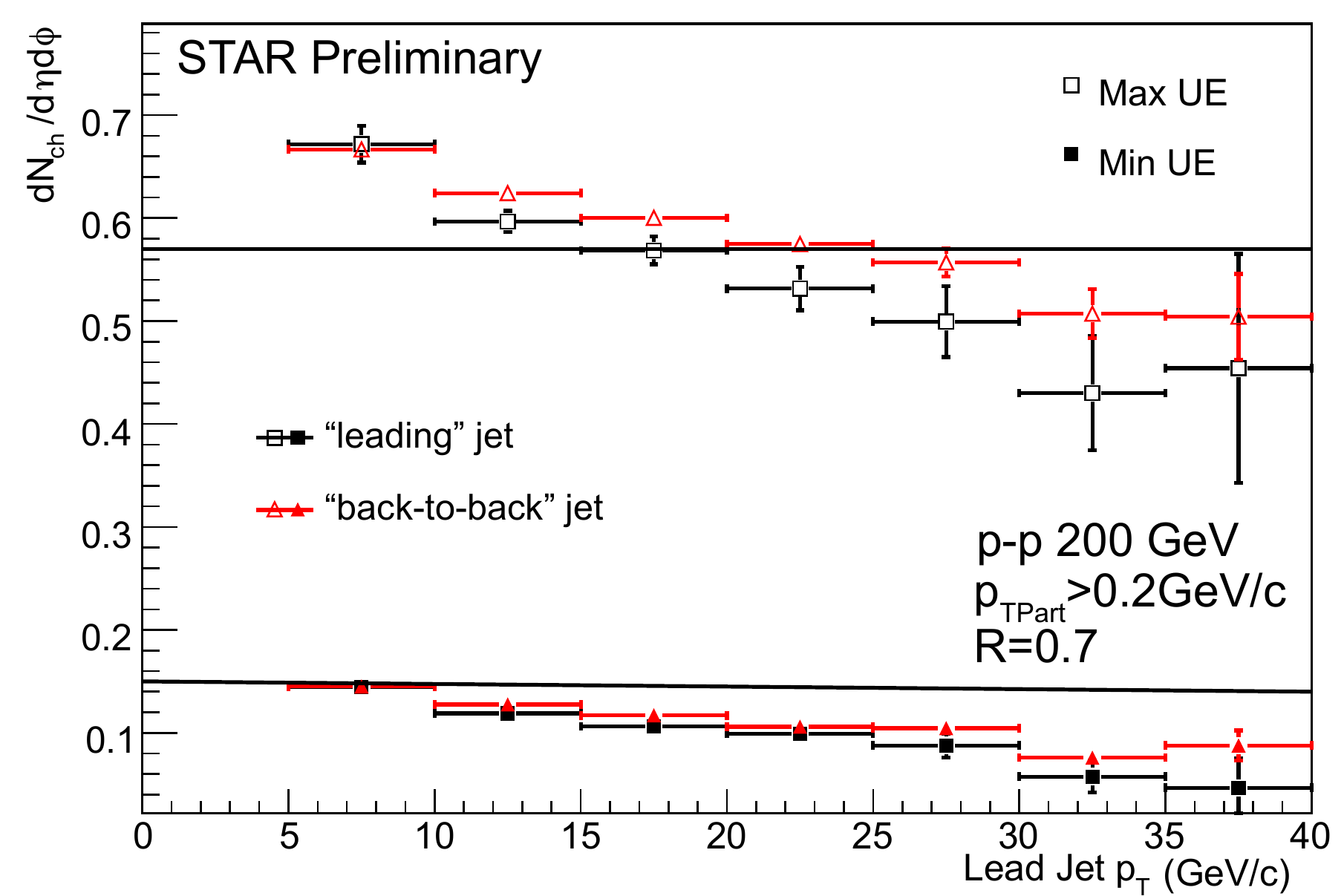}
			\caption{Color online: The uncorrected charged particle density in the TransMin and TransMax regions as a function of reconstructed lead jet \pT, using SISCone algorithm, R=0.7.  }
			\label{Fig:UE}
		\end{center}

\end{figure}

Figure~\ref{Fig:UE} shows the measured charged particle density in the UE. The first observation is that the UE is largely independent of the jet energy. The second is that the densities are the same within errors for the ``leading" and ``back-to-back" datasets. This again suggests that the hard scattered partons emit  very small amounts of large angle initial/final state radiation at RHIC energies. This is very different in 1.96 TeV collisions where the ``leading"/``back-to-back" density ratio is $\sim$0.65~\cite{CDF}.  The two solid lines show the expected density if  events follow a Poisson distribution with an average of 0.36. The similarity of this simple simulation to the data suggests that at RHIC energies the splitting of the measured TransMax and TransMin values  is predominantly due to the sampling. PYTHIA again shows satisfactory agreement with the data.

\section{Summary}

In summary, jet fragmentation functions have been measured in \pp collisions at \sqrts  = 200 GeV and will provide a stringent baseline for the measurements underway in Au-Au collisions. PYTHIA, tuned to 1.96 TeV data, shows reasonable agreement suggesting that the energy dependence of the underlying physics is well modeled.
Finally, the UE is largely independent of the momentum transfer of hard scattering and receives only minor contributions from radiation from this scattering.

\end{document}